\documentclass[a4paper]{panl}
\usepackage{cite}
\usepackage{wrapfig}
\usepackage{graphicx}
\usepackage{amssymb}
\usepackage{amsfonts}
\usepackage{amsmath}
\usepackage{longtable}
\usepackage{rotating}
\usepackage{lscape}
\usepackage{epsfig}
\usepackage{multirow}

\usepackage{bm}
\usepackage{gensymb}
\usepackage{slashed}
\usepackage{hyperref}

\newcommand{\ltsim}{\protect\raisebox{-0.5ex}{$\:\stackrel{\textstyle <}
	{\sim}\:$}}
\newcommand{\gtsim}{\protect\raisebox{-0.5ex}{$\:\stackrel{\textstyle >}
	{\sim}\:$}}
\newcommand{\bvec}[1]{\ensuremath{\boldsymbol{#1}}}

\russianTeX
\begin{document}

\title{Color transparency in hard $pd$ collisions \\ Цветовая прозрачность в жестких $pd$ столкновениях}
\maketitle
\authors{A.B.\,Larionov \footnote{E-mail: larionov@theor.jinr.ru}}
\setcounter{footnote}{0}
\authors{А.Б.\,Ларионов}

\from{Bogoliubov Laboratory of Theoretical Physics, Joint Institute for Nuclear Research, 141980 Dubna, Russia}
\from{Лаборатория теоретической физики им. Н.Н. Боголюбова, ОИЯИ, 141980 Дубна, Московская область, Россия}

\begin{abstract}
  Являясь одним из предсказаний пертурбативной КХД, эффект цветовой прозрачности находится в центре внимания
  сообщества, изучающего модификации адронов в ядерной среде, уже несколько десятилетий. Поиск этого эффекта
  в реакциях на тяжелых ядрах может быть затруднен неопределенностями характеристик ядра
  (распределения плотности нуклонов и волновые функции), которые могут повлиять на интерпретацию экспериментов.
  В настоящей работе рассматривается реакция $d(p,pp)n$ при $p_{\rm lab}=15$ ГэВ/c, вызванная жестким упругим $pp$ рассеянием,
  в которой данные неопределенности фактически сводятся к поведению волновой функции дейтрона при  больших импульсах.
  Показано, что при поперечных импульсах нейтрона-спектатора $\ltsim 0.4$ ГэВ/c выбор волновой функции дейтрона не может
  повлиять на идентификацию эффекта цветовой прозрачности. Предлагается также простой способ изучения цветовой прозрачности
  в $dd$ столкновениях на основе выделения квазисвободного $pd$ взаимодействия.

\vspace{0.2cm}

As one of the predictions of perturbative QCD, the effect of color transparency has been the focus of attention
in the community studying modifications of hadrons in nuclear medium for several decades.
The search for this effect in reactions involving heavy nuclei can be complicated by uncertainties in nuclear characteristics
(nucleon density distributions and wave functions), which can affect the interpretation of experiments.
In this work, we consider the reaction $d(p,pp)n$ at $p_{\rm lab}=15$ GeV/c caused by hard elastic $pp$ scattering,
in which these uncertainties are actually reduced to the behavior of the deuteron wave function at large momenta.
It is shown that for transverse momenta of the spectator neutron $\ltsim 0.4$ GeV/c
the choice of the deuteron wave function cannot affect the identification of the color transparency effect.
A simple method for studying color transparency in $dd$ collisions is also suggested
based on the identification of quasi-free $pd$ interactions.
\end{abstract}
\vspace*{6pt}

\noindent
PACS: 24.10.Ht; 25.40.-h; 25.45.-z; 24.50.+g; 24.70.+s

\section*{Введение}
\label{sec:intro}

Как известно, жесткие процессы на нуклоне характеризуются большими передачами импульса, $Q^2 \gg 1$ GeV$^2$. Для их описания необходимо вводить
кварк-глюонные степени свободы. При этом поперечный размер нейтральных по цвету кварковых ($q\bar q$ или $qqq$) конфигураций в начальном и/или конечном состоянии можно оценить
как $r_\perp \sim 1/Q$. В пределе больших $Q$ кварковые конфигурации становятся точечно-подобными. 
В рамках пертурбативной КХД сечение взаимодействия цвето-нейтральных точечно-подобных $q\bar q$ конфигураций с нуклоном можно оценить как $\sigma_{q\bar q} \propto r_\perp^2 \sim 1/Q^2$,
т.е. сечение ведёт себя чисто геометрически (см., например, уравнение (1) в обзоре \cite{Dutta:2012ii}).
\footnote{То же самое разумно предположить и относительно сечения взаимодействия точечно-подобных $qqq$ конфигураций с нуклоном, хотя строгое доказательство здесь отсутствует.}
Такое поведение сечений называют эффектом цветовой прозрачности (color transparency, CT): цвето-нейтральная кварковая конфигурация в начальном или конечном состонии эксклюзивного процесса
с большой передачей импульса взаимодействует с нуклонами с уменьшенным сечением \cite{Dutta:2012ii}.

Наблюдаемой, чувствительной к эффекту CT, является ядерная прозрачность, т.е. отношение измеренного сечения для определенного жесткого процесса на ядре
в определенной кинематике к тому же сечению, рассчитанному в импульсном приближении (impulse approximation, IA) 
\begin{equation}
      T=\frac{\sigma}{\sigma_{\rm IA}}~.       \label{T}
\end{equation}
Для инклюзивных по ядру-остатку процессов, инициированных взаимодействием с протонами ядра, в пренебрежении ядерным ферми-движением получаем
$\sigma_{\rm IA} \simeq Z \sigma_p$, где $\sigma_p$ -- сечение на протоне. Без учета перерассеяния частиц преобладает ядерное поглощение,
что дает в простейшем глауберовском приближении $T < 1$. Режим полной CT означает $T=1$.

В полной мере эффект CT ожидается при  ультрарелятивистких энергиях (энергии пучка $\gtsim$ 100 ГэВ).
В области промежуточных энергий (энергии пучка $\sim$ 10 ГэВ)  наблюдение CT осложняется расширением точечно-подобных конфигураций до нормального адронного размера по мере прохождения
через ядро-остаток. Поэтому здесь ожидается рост $T$ с энергией столкновения от ``стандартных'' глауберовских величин до $T \sim 1$. В эксперименте EVA@AGS \cite{Leksanov:2001ui}
исследовалась ядерная прозрачность в реакции квазиупругого выбивания протона $^{12}\mbox{C}(p,pp)$ при $\Theta_{c.m.}=90\degree$. Оказалось, что величина $T$ растет с энергией
протонного пучка от 6 до 9 ГэВ в согласии с CT. Однако в диапазоне от 9 до 14.5 ГэВ  $T$ падает вплоть до глауберовских значений. Предложенные в литературе объяснения данного спада
основаны на введении либо  интерференции кварковых конфигураций различного размера \cite{Ralston:1988rb}, либо промежуточного $6qc\bar c$ резонанса с массой $\sim 5$ ГэВ
и шириной $\sim 1$ ГэВ \cite{Brodsky:1987xw}.    

В работе \cite{Frankfurt:1996uz} было предложено исследовать CT в реакции $d(p,2p)n$. Помимо относительной простоты волновой функции, другое преимущество дейтрона состоит в том, что
изменяя поперечный импульс спектаторного нейтрона можно регулировать влияние взаимодействий в начальном и конечном состоянии на амплитуду процесса, что даёт
больший контроль над эффектом CT по сравнению с реакциями на более тяжелых ядерных мишенях, где ядерный остаток обычно не детектируется.  
Авторы  \cite{Frankfurt:1996uz} сформулировали метод обобщенного эйконального приближения (generalized eikonal approximation, GEA) и предложили способ введения эффекта CT
в рамках модели квантовой диффузии. В \cite{Frankfurt:1996uz} расчеты ядерной прозрачности  были проведены при $p_{\rm lab}=6-18$ ГэВ/c ($\sqrt{s_{NN}}=3.63-5.96$ ГэВ).

В работе \cite{Larionov:2022gvn} модель \cite{Frankfurt:1996uz} была дополнена введением интерференции кварковых конфигураций согласно \cite{Ralston:1988rb},
и реакция $d(p,2p)n$ была рассмотрена в более широком диапазоне $p_{\rm lab}=6-75$ ГэВ/c ($\sqrt{s_{NN}}=3.63-11.94$ ГэВ),
что соответствует энергетическому диапазону NICA SPD  \cite{Arbuzov:2020cqg,Abramov:2021vtu,SPD:2024gkq}. 
В \cite{Larionov:2022gvn} были рассчитаны дифференциальные сечения, ядерная прозрачность и тензорная анализирующая способность дейтрона.
При этом оказалось, что эти величины наиболее чувствительны к эффектам CT при довольно больших поперечных импульсах спектаторного нейтрона, $p_{\rm st} = 0.2-0.4$ GeV/c,
где становятся существенными вклады диаграмм с перерассеянием в начальном и конечном состоянии. В то же время ясно, что с увеличением поперечного импульса нейтрона
растет также и неопределенность волновой функции дейтрона (ВФД). 

В настоящей работе решались две задачи. (i) Исследование влияния выбора ВФД на наблюдаемые для реакции жесткого выбивания протона, $d(p,2p)n$, при энергиях NICA SPD.    
(ii) Обобщение модели на случай $dd$ столкновений, которые более предпочтительны на первой фазе эксперимента NICA SPD \cite{SPD:2024gkq}.

\section*{Чувствительность к выбору волновой функции дейтрона}
\label{sec:DWF}

\begin{figure}[t]
  \begin{center}
  \includegraphics[scale = 0.50]{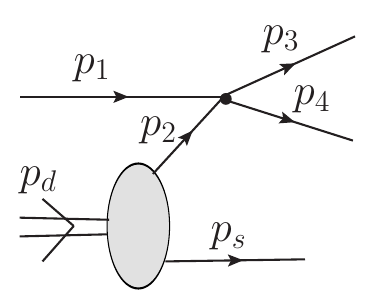}
  \vspace{-3mm}
  \caption{\label{fig:IA} Диаграмма IA для процесса $p d  \to p p n$.
    Четырехимпульсы дейтрона, нейтрона-спектатора, протона пучка, промежуточного протона, и вылетающих протонов обозначены
    как $p_d$, $p_s$, $p_1$, $p_2$, $p_3$, $p_4$, соответственно.
    Эллипсоид обозначает вершину виртуального распада дейтрона на нейтрон в конечном состоянии и виртуальный протон, $\Gamma_{d \to pn}(p_d,p_s)$.
    Точка обозначает величину $iM_{\rm hard}$, где $M_{\rm hard}(s,t)$ -- инвариантная амплитуда жесткого процесса $pp \to pp$,
    $s=(p_3+p_4)^2$, $t=(p_1-p_3)^2$.}
  \end{center}
\end{figure}

Модель расчета детально описана в \cite{Frankfurt:1996uz,Larionov:2022gvn}.
В основе лежит инвариантная амплитуда IA, Рис.~\ref{fig:IA},
которая записывается в виде
\begin{equation}
    M_{pd}^{\rm IA} =  \frac{i\Gamma_{d \to pn}(p_d,p_s)}{p_2^2-m^2+i\epsilon} M_{\rm hard}(s,t)
    = 2m^{1/2} (2\pi)^{3/2} \phi^{\lambda_d}(-\bvec{p}_s^r) M_{\rm hard}(s,t)~,    \label{M^(a)_fin}
\end{equation}
где $m$ -- масса нуклона; $\phi^{\lambda_d}(-\bvec{p}_s^r)$ -- ВФД, определенная в системе покоя дейтрона,
что обозначено верхним индексом ``r'' у импульса нейтрона-спектатора; $\lambda_d$ -- проекция спина дейтрона.
Второй шаг в (\ref{M^(a)_fin}) справедлив, когда четырехимпульс нейтрона-спектатора находится на массовой поверхности, т.е. $p_s^2=m^2$.

В добавление к амплитуде IA в модели когерентно учтены также амплитуды с однократным и двукратным перерассеянием на нейтроне-спектаторе, которые вычислялись в GEA.
Эффекты CT вводились в амплитуды с перерассеянием в модели квантовой диффузии  \cite{Farrar:1988me}.
В данной модели предполагается, что сечение взаимодействия быстрых адронов, входящих в точку жесткого взаимодействия или выходящих из нее, с нуклонами ядра-остатка
линейно увеличивается с расстоянием от этой точки от малой величины $\propto |t|^{-1}$ до нормального адронного сечения.
Расстояние, на котором достигается нормальное адронное сечение называется длиной когерентности
\footnote{Название связано с тем, что на той же длине теряется когерентность между состояниями с заданными импульсом $\bvec{p}$ и
квадратами масс, отличающимися на $\Delta M^2$. См. разд. 3.1 в работе \cite{Dutta:2012ii}.} и линейно зависит от импульса адрона:
\begin{equation}
  l_c = \frac{2p}{\Delta M^2}~.   \label{l_c}
\end{equation}
Значение $\Delta M^2$ является параметром модели квантовой диффузии, от которого зависит подавление взаимодействий в начальном и конечном состоянии за счет CT
и который выбирается из согласия с экспериментом. Выбор $\Delta M^2 \simeq 1$ ГэВ$^2$ предпочтителен для описания данных JLab по ядерной прозрачности
в реакциях  $A(e,e^\prime \pi)$ \cite{Larson:2006ge}. Тот же выбор согласуется с анализом данных AGS по ядерной прозрачности в реакциях $A(p,2p)$
в рамках модели RMSGA \cite{VanOvermeire:2006tk}. Однако недавние данные JLab для реакции $^{12}$C$(e,e^\prime p)$  \cite{HallC:2020ijh}
требуют $\Delta M^2 = 2-3$ ГэВ$^2$, как показывают RMSGA расчеты \cite{Li:2022uvf}. В настоящей работе, чтобы описать неопределенность эффекта CT,
выбирались значения $\Delta M^2=0.7$ и 3 ГэВ$^2$.

Как и в предыдущих расчетах \cite{Frankfurt:1996uz,Larionov:2022gvn}, по умолчанию будем использовать ВФД в модели с парижским потенциалом \cite{Lacombe:1981eg} ($P_D=5.8\%$).
Чтобы исследовать чувствительность к ВФД, воспользуемся ВФД в модели CD-Бонн \cite{Machleidt:2000ge} ($P_D=4.9\%$). Здесь в скобках указана фракция D-волновой компоненты.

Расчет проводился в системе покоя дейтрона c осью $z$ вдоль импульса налетающего протона. 
Импульс протона выбирался равным 15 ГэВ/c ($\sqrt{s_{NN}}=5.47$ ГэВ). 
Кинематика выходного канала определялась следующими переменными: 
\begin{equation}
  \alpha_s = \frac{2(p_s^0-p_s^z)}{m_d}                 \label{alpha_s}
\end{equation}
-- переменная светового конуса, $p_{st}$ -- поперечный импульс нейтрона, $t$,
$\phi=\phi_3-\phi_s$ -- относительный азимутальный угол между импульсом вылетающего протона и нейтрона.
Во всех расчетах полагалось $\alpha_s=1$, что при малых $p_{st}$ соответствует поперечному направлению вылета нейтрона,
и $t=(4m^2-s)/2$, что соответствует углу жесткого упругого $pp$ рассеяния в системе центра масс (с.ц.м.) $\Theta_{c.m.}=90\degree$.     

\begin{figure}[t]
  \begin{center}
    \includegraphics[scale = 0.40]{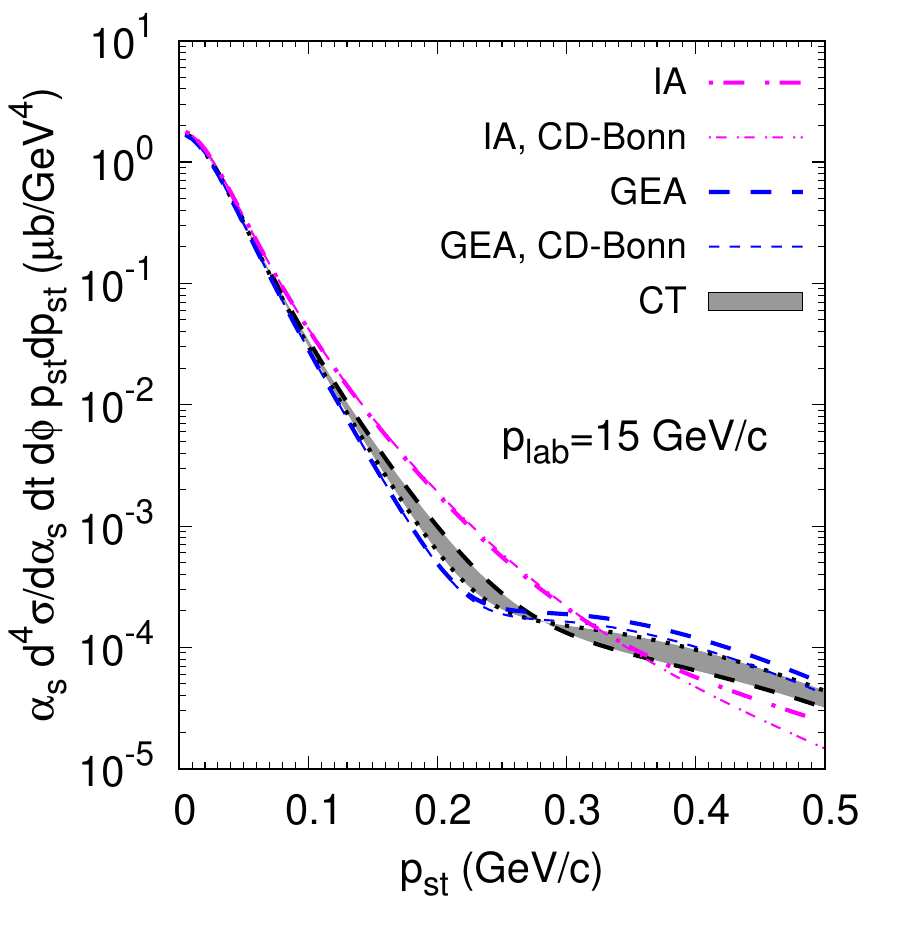}
    \vspace{-3mm}
    \caption{\label{fig:sig_15gevc_180deg} Зависимость дифференциального сечения $p d  \to p p n$ от поперечного импульса нейтрона при $\phi=180\degree$.
      Расчеты в IA и GEA показаны штрихпунктирными пурпурными и штриховыми синими линиями, соответственно.
      Толстые и тонкие линии соответствуют расчетам с парижским и боннским потенциалами.
      Серая  полоса показывает расчет с включением эффекта CT для случая парижского потенциала. Полоса ограничена черными штриховой и пунктирной линиями,
      которые соответствуют значениям параметра длины когерентности $\Delta M^2=0.7$ и 3 ГэВ$^2$.}  
  \end{center}
\end{figure}
На Рис.~\ref{fig:sig_15gevc_180deg} показано дифференциальное сечение как функция $p_{st}$. При малых поперечных импульсах интерференция амплитуд однократного
перерассеяния с амплитудой IA приводит к уменьшению сечения по сравнению с расчетом в IA, что отвечает глауберовскому поглощению. При больших поперечных импульсах доминирует квадрат
модуля суммы амплитуд перерассеяния, дающий увеличение сечения по сравнению с IA. Учет CT смещает результаты в сторону IA.   
Как видно, в IA расчете использование ВФД CD-Бонн даёт заметное уменьшение сечения при $p_{st} > 0.4$ ГэВ/c. Однако в полном GEA расчете разница между сечениями значительно меньше.
Это объясняется тем, что амплитуды с перерассеянием определяются главным образом низкоимпульсной частью ВФД, которая мало зависит от используемого потенциала.

\begin{figure}[t]
  \begin{center}
    \includegraphics[scale = 0.40]{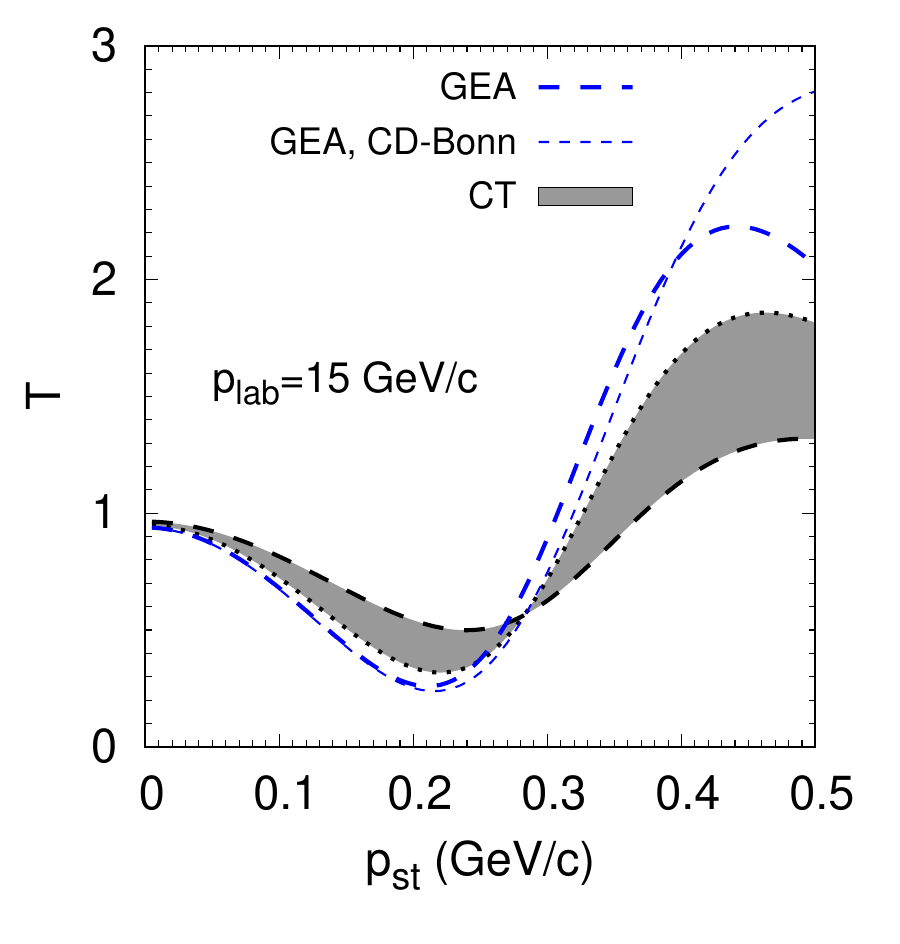}
    \vspace{-3mm}
    \caption{\label{fig:T_15gevc_180deg} То же, что и на Рис.~\ref{fig:sig_15gevc_180deg}, но для ядерной прозрачнности.}  
  \end{center}
\end{figure}
Рисунок~\ref{fig:T_15gevc_180deg} показывает зависимость прозрачности $T$ от поперечного импульса нейтрона. По определению, в IA $T=1$.
При $p_{st} < 0.2$ ГэВ/c, т.е. в области глауберовского поглощения, прозрачность падает с увеличением $p_{st}$. При $p_{st} > 0.2$ ГэВ/c прозрачность начинает расти
с увеличением $p_{st}$, достигая значений больше единицы при $p_{st} \gtsim 0.3$ ГэВ/c, где доминируют амплитуды с перерассеянием на спектаторе.
Включение эффекта CT приводит к смещению $T$ ближе к единице. 
При $p_{st} > 0.4$ ГэВ/c $T$ довольно сильно зависит от выбора ВФД, что связано в основном с чувствительностью дифференциального сечения в IA к выбору ВФД .

Рассмотрим теперь тензорную анализирующую способность дейтрона. Согласно \cite{Platonova:2010wjt,Uzikov:2020zho} она определяется как
\begin{equation}
   A_{\alpha\beta} = \frac{\mbox{Sp}(MS_{\alpha\beta}M^\dag)}{\mbox{Sp}(MM^\dag)}~,~~~\alpha,\beta=x,y,z,    \label{A_alpha_beta}
\end{equation}
где $M$ -- амплитуда процесса, $S_{\alpha\beta} = \frac{3}{2}(S_\alpha S_\beta  + S_\beta S_\alpha) - 2\delta_{\alpha\beta}$ -- спин-квадрупольный оператор,
$\bvec{S}$ -- оператор спина дейтрона. Диагональные компоненты тензора (\ref{A_alpha_beta}) могут приведены к следующему виду:
\begin{equation}
    A_{\alpha\alpha} = \frac{\sigma_\alpha(+1)+\sigma_\alpha(-1)-2\sigma_\alpha(0)}{\sigma_\alpha(+1)+\sigma_\alpha(-1)+\sigma_\alpha(0)}~,     \label{A_alpha_alpha}
\end{equation}
где $\sigma_\alpha(\lambda_d)$ -- дифференциальное сечение при фиксированной проекции спина дейтрона $\lambda_d$ на ось $\alpha$.
В IA при независящей от спина амплитуде жесткого рассеяния вместо сечений в формулу (\ref{A_alpha_alpha})
можно подставить квадраты модуля ВФД с соответствующими проекциями спина, что дает:
\begin{equation}
  A_{\alpha\alpha}^{IA} = \frac{(3(p_s^\alpha/p_s)^2 -1)(\sqrt{2}u(p_s)w(p_s) - w^2(p_s)/2)}{u^2(p_s) +w^2(p_s)}~.       \label{A_alpha_alpha^IA}
\end{equation}
Здесь $u(p_s)$ и $w(p_s)$ -- S- и D-волновые компоненты ВФД, соответственно. 
Таким образом, в IA тензорная анализирующая способность чувствительна к D-компоненте ВФД.
В настоящей работе спин дейтрона квантовался относительно оси $z$ (продольная поляризация). 

\begin{figure}[t]
  \begin{center}
    \includegraphics[scale = 0.40]{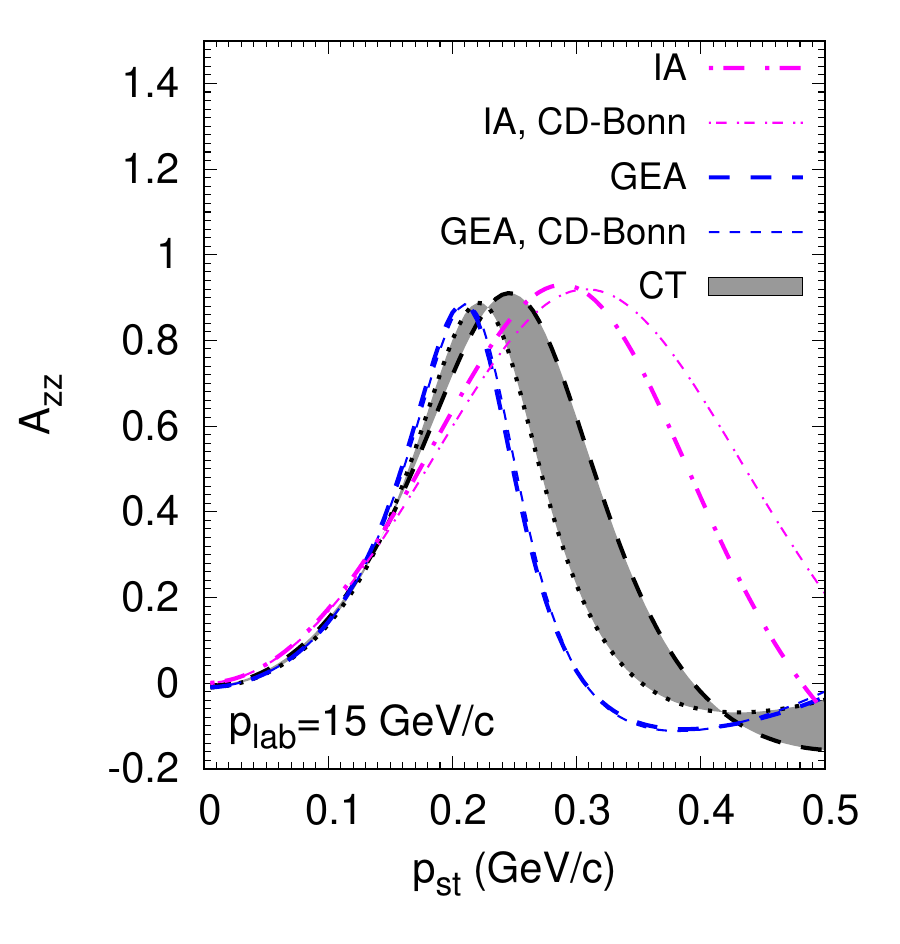}
    \vspace{-3mm}
    \caption{\label{fig:Azz_15gevc_180deg} То же, что и на Рис.~\ref{fig:sig_15gevc_180deg}, но для тензорной анализирующей способности дейтрона.}  
  \end{center}
\end{figure}
На Рис.~\ref{fig:Azz_15gevc_180deg} показана  зависимость компоненты $A_{zz}$ от поперечного импульса нейтрона.
В IA $A_{zz}$ имеет широкий максимум вблизи $p_{st}=0.3$ ГэВ/c. При этом видна сильная зависимость от ВФД при больших поперечных импульсах нейтрона.
В полном GEA расчете максимум $A_{zz}$  сдвигается к $p_{st}=0.2$ ГэВ/c и сужается, а зависимость от ВФД практически исчезает.
Эффект включения CT наиболее заметен вблизи $p_{st}=0.3$ ГэВ/c.

\section*{$dd$ столкновения}
\label{sec:dd}

Дейтрон является слабосвязанным ядром со среднеквадратическим расстоянием между протоном и нейтроном $\simeq 4$ Фм. Поэтому при взаимодействии с другим ядром
протон и нейтрон дейтрона с большой вероятностью ведут себя как квазисвободные частицы. В случае столкновения двух дейтронов можно потребовать, чтобы
нейтрон-спектатор одного из дейтронов имел малый поперечный импульс ($\ltsim 0.1$ ГэВ/c), что практически исключит возможные перерассеяния других нуклонов на этом нейтроне
и сведёт $dd$ столкновение к $pd$ столкновению.  
\begin{figure}[t]
  \begin{center}
    \includegraphics[scale = 0.50]{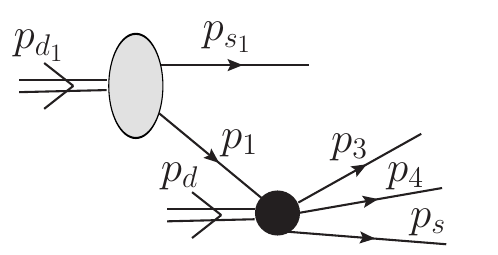}
    \vspace{-3mm}
    \caption{\label{fig:dd} Фейнмановская диаграмма процесса $dd \to ppnn$. Линии обозначены четырехимпульсами частиц: начальные дейтроны ($p_d$ и $p_{d_1}$),
      нейтроны-спектаторы ($p_s$ и $p_{s_1}$), промежуточный протон ($p_1$), конечные протоны ($p_3$ и $p_4$).
      Точка обозначает величину $iM_{pd}$, где $M_{pd}(p_3,p_4,p_s;p_1)$ -- инвариантная амплитуда процесса $pd \to ppn$.}
  \end{center}
\end{figure}
На Рис.~\ref{fig:dd} показана диаграмма процесса $dd \to ppnn$, в котором предполагается, что поперечный импульс $p_{s_1,t}$ мал.
Соответствующая инвариантная амплитуда записывается аналогично (\ref{M^(a)_fin}) как
\begin{equation}
   M_{dd} = 2m^{1/2} (2\pi)^{3/2} \phi^{\lambda_1}(-\bvec{p}_{s_1}^r) M_{pd}(p_3,p_4,p_s;p_1)~,        \label{M_dd}
\end{equation}
Квадрат модуля амплитуды (\ref{M_dd}) имеет вид:
\begin{equation}
    \overline{|M_{dd}|^2} = 4m (2\pi)^{3} \overline{|\phi(-\bvec{p}_{s_1}^r)|^2}\,\overline{|M_{pd}|^2}~,     \label{M_dd^2}
\end{equation}
где черта сверху обозначает усреднение по проекциям спина начальных частиц и суммирование по проекциям спина конечных частиц.
\footnote{При вычислении поляризационных наблюдаемых усреднение по проекции спина дейтрона $d_1$ на Рис.~\ref{fig:dd} не проводится.}
Усредненный по проекции спина дейтрона квадрат модуля ВФД выражается через ее S- и D-компоненты:
\begin{equation}
  \overline{|\phi(\bvec{p})|^2} = \frac{1}{3} \sum_{\lambda_1} |\phi^{\lambda_1}(\bvec{p})|^2
                                     = \frac{u^2(p)+w^2(p)}{4\pi}           \label{phi^2}
\end{equation}
с условием нормировки
\begin{equation}
   \int dp p^2 (u^2(p)+w^2(p)) = 1~.     \label{norm}
\end{equation}
Полное дифференциальное сечение процесса $dd \to ppnn$ записывается в стандартном виде:
\begin{eqnarray}
  d\sigma &=& (2\pi)^4 \delta^{(4)}(p_{s_1}+p_3+p_4+p_s-p_{d_1}-p_d) \frac{\overline{|M_{dd}|^2}}{4I_{dd}} \nonumber \\
          && \times \frac{d^3p_{s_1}}{(2\pi)^32E_{s_1}} \frac{d^3p_3}{(2\pi)^32E_3}  \frac{d^3p_4}{(2\pi)^32E_4} \frac{d^3p_s}{(2\pi)^32E_s}~,
     \label{dsigma}
\end{eqnarray}
где $I_{dd}=[(p_{d_1}p_d)^2 - m_d^4]^{1/2}$ -- фактор потока, $m_d$ -- масса дейтрона. Подставляя (\ref{M_dd^2}) в (\ref{dsigma}) приходим к выражению:
\begin{equation}
    d\sigma \simeq  \overline{|\phi(-\bvec{p}_{s_1}^r)|^2}\, d^3p_{s_1}^r\, d\sigma_{1d \to 34s}~,   \label{dsigma_fact}
\end{equation}
где
\begin{eqnarray}
    d\sigma_{1d \to 34s} &=& (2\pi)^4 \delta^{(4)}(p_3+p_4+p_s-p_1-p_d) \frac{\overline{|M_{pd}|^2}}{4I_{pd}} \nonumber \\
                      && \times \frac{d^3p_3}{(2\pi)^32E_3}  \frac{d^3p_4}{(2\pi)^32E_4} \frac{d^3p_s}{(2\pi)^32E_s}   
    \label{dsigma_1d}
\end{eqnarray}
-- полное дифференциальное сечение взаимодействия квазисвободного протона с дейтроном,
$I_{pd}=[(p_1p_d)^2 - m^2m_d^2]^{1/2}$ -- соответствующий фактор потока. При получении выражения (\ref{dsigma_fact})
было использовано условие квазисвободности протона ($p_1 \simeq p_{d_1}/2$), а также лоренц-инвариантность отношения $d^3p_{s_1}/E_{s_1}$. 
Формула (\ref{dsigma_fact}) позволяет вычислить любое дифференциальное сечение процесса  $dd \to ppnn$ исходя из
сечения соответствующего квазисвободного процесса  $pd \to ppn$.

\section*{Заключение}
\label{sec:concl}

В рамках модели GEA с учетом эффектов CT \cite{Larionov:2022gvn} проведены дополнительные расчеты процесса $pd \to ppn$,
вызванного жестким рассеянием $pp \to pp$ на большие углы в с.ц.м. Целью являлось выяснение неопределенности результатов,
полученных в \cite{Larionov:2022gvn}, к выбору ВФД. Показано, что при поперечном импульсе нейтрона-спектатора $p_{st} < 0.4$ ГэВ/c
выбор ВФД мало влияет на ядерную прозрачость $T$, однако при $p_{st} > 0.4$ ГэВ/c влияние ВФД на $T$ становится сопоставимым с влиянием эффекта CT.
Тензорная анализирующая способность $A_{zz}$ оказалась практически не чувствительной к выбору ВФД при $p_{st} < 0.5$ ГэВ/c, что делает эту
наблюдаемую предпочтительной для выявления эффектов CT при больших поперечных импульсах нейтрона-спектатора.

Модель обобщена на случай процесса $dd \to ppnn$, также вызванного жестким рассеянием $pp \to pp$,  при условии, что один из
нейтронов-спектаторов имеет малый поперечный импульс.
\footnote{В общем случае произвольных поперечных импульсов нейтронов-спектаторов требуется учет дополнительных процессов перерассеяния, что усложняет
расчет и выходит за рамки настоящей работы.}
Это открывает возможность исследования CT в процессе $dd \to ppnn$.

\bibliographystyle{pepan}
\bibliography{pd2ppn_ran24}

\end{document}